# A New Amplification Regime for Traveling Wave Tubes with Third Order Modal Degeneracy

Farshad Yazdi, Mohamed A. K. Othman, Mehdi Veysi, Alexander Figotin, and Filippo Capolino

*Abstract*— Engineering of the eigenmode dispersion of slow-wave structures (SWSs) to achieve desired modal characteristics, is an effective approach to enhance the performance of high power traveling wave tube (TWT) amplifiers or oscillators. We investigate here for the first time a new synchronization regime in TWTs based on SWSs operating near a third order degeneracy condition in their dispersion. This special three-eigenmode synchronization is associated with a stationary inflection point (SIP) that is manifested by the coalescence of three Floquet-Bloch eigenmodes in the SWS. We demonstrate the special features of "cold" (without electron beam) periodic SWSs with SIP modeled as coupled transmission lines (CTLs) and investigate resonances of SWSs of finite length. We also show that by tuning parameters of a periodic SWS one can achieve an SIP with nearly ideal flat dispersion relationship with zero group velocity or a slightly slanted one with a very small (positive or negative) group velocity leading to different operating schemes. When the SIP structure is synchronized with the electron beam potential benefits for amplification include (i) gain enhancement, (ii) gain-bandwidth product improvement, and (iii) higher power efficiency, when compared to conventional Pierce-like TWTs. The proposed theory paves the way for a new approach for potential improvements in gain, power efficiency and gain-bandwidth product in high power microwave amplifiers.

## I. Introduction

The classical approach for designing high power microwave amplifiers provides for an efficient energy transfer from high energy electron beams to electromagnetic fields at radio and microwave frequencies [1], [2]. A traveling wave tube (TWT) amplifier is a conventional high power device comprising of a slow-wave structure (SWS) whose interacting mode has a synchronous phase velocity to the average electron's velocity of the electron beam [1]–[4]. Pierce and his contemporaries [3], [5]–[8] developed a ubiquitous framework and design procedure for such devices through circuit theory. According to simple but physically incisive Pierce model [3], the amplification in a TWT is attributed to amplification of a slow wave radio frequency (RF) signal in an equivalent transmission line (TL) due to perturbation of the electron charge density thanks to bunching of the electron beam (the charge wave). Remarkably, Pierce predicted the small signal gain of a TWT and provided design rules for TWT amplifiers in terms of the SWS and electron beam parameters [2], [9]. In essence, state of the art of the high power TWT technology employs all-metallic slow-wave guiding structures whose dispersion is engineered for (i) matching phase velocity to an electron beam over a wide bandwidth; and (ii) high interaction impedance [10]–[17]. Consequently, dispersion engineering of the SWS eigenmodes would potentially enhance the gain, efficiency and bandwidth of conventional TWTs [18]–[22] and backward wave oscillators (BWOs) [23]–[26]. We investigate here a novel amplification regime based on special dispersion characteristics of SWS potentially leading to a higher gain and larger gain-bandwidth product (typical figure of merit for amplifiers) compared to a conventional Pierce-type TWT. In particular, the proposed regime of operation relies on electromagnetic eigenmode degeneracy in periodic SWS, namely, the third order degeneracy typically referred to as the stationary inflection points (SIP). The SIP condition is found when three Floquet-Bloch eigenmodes in the "cold" periodic structure coalesce [27]–[30] and cause an inflection point in the dispersion diagram. The "cold" term refers to a SWS which is not coupled to an electron beam. In [31]–[33], some of the authors have developed the theory of a SWS-electron beam interaction based on a different modal degeneracy, the degenerate band edge (DBE) [34]–[38], which inherently has limited bandwidth. The theory in [31]–[33], [39] describes four Floquet-Bloch eigenmodes synchronous scheme of oscillators operating with a fourth order degeneracy at the degenerate band edge (DBE).

In this premise, instead, we investigate amplification in TWTs operating near the SIP in their dispersion relation. Such operational principle is referred to as *three Floquet-Bloch eigenmode interaction* which promises special features which are not present in conventional single mode TWTs, or even in TWTs operating at the DBE. The main reason for these special features is that a SWS with SIP exploits three degenerate eigenmodes that *(i) do not have negative group velocity*, and *(ii)*

This material is based upon work supported by the Air Force Office of Scientific under the Multidisciplinary University Research Initiative award number FA9550-12-1-0489 administered through the University of New Mexico, and under award number FA9550-15-1-0280.

F. Yazdi, M. A. K. Othman, M. Veysi, and F. Capolino are with the Department of Electrical Engineering and Computer Science, University of California, Irvine, CA 92697 USA. (e-mail: fyazdi@uci.edu, mothman@uci.edu, mveysi@uci.edu, f.capolino@uci.edu).

A. Figotin is with the Department of Mathematics, University of California, Irvine, CA 92697 USA. (e-mail: afigotin@uci.edu, a.figotin@uci.edu).

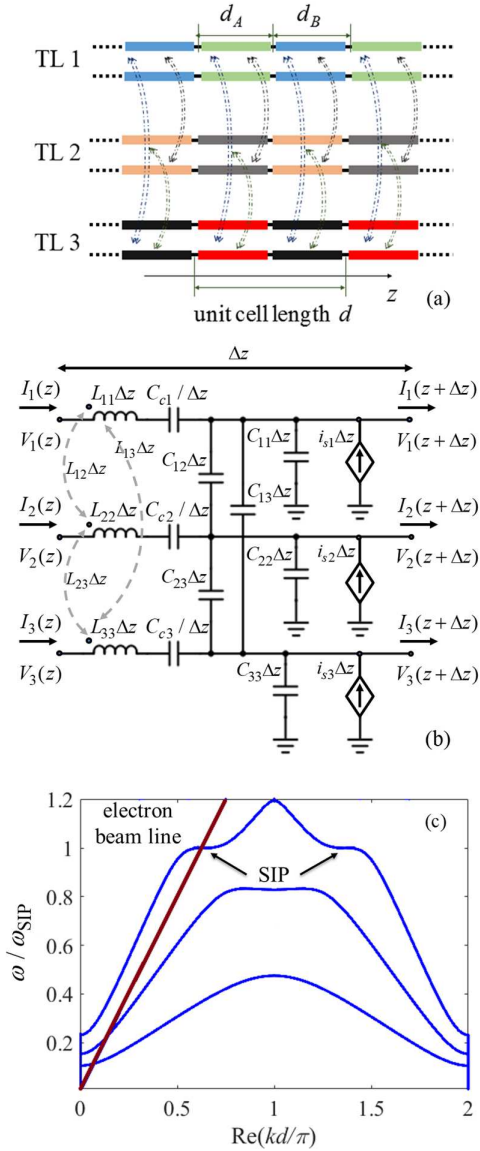

Fig. 1 (a) Schematic of periodic CTLs consisting of three TLs that mimic a SWS supporting three coupled eigenmodes. The unit cell, of period $d$, is made of two segments (shown by different colors). Each segment is described using TLs distributed parameters. (b) Generalized Pierce model for the three CTL interacting with an electron beam. The dependent current sources are used in the generalized Pierce model to describe the interaction of the charge wave modulating the electron beam with the RF signal in the SWS. (c) Typical modal dispersion diagram, where only branches with purely real Floquet-Bloch wavenumbers are shown, of a periodic "cold" three CTLs (without including the beam interaction) having SIP at angular frequency $\omega_{\mathrm{SIP}}$. The beam line shows the charge wave velocity to be synchronized with the three RF modes in the TLs that is introduced in (2) and discussed in Sec. IV. Dispersion diagrams for the combined CTL-electron beam interactive system are shown in Sec. IV-A and Appendix A.

have frequencies inside the band and are far from any bandgap. In contrast, an SWS based on the DBE exploits both positive and negative group velocity in the DBE neighborhood that has been proposed for a new regime of operation for low-threshold oscillators [33].

We demonstrate (i) a significant gain enhancement of the proposed TWT with SIP in comparison to a periodic single mode Pierce one (orders of magnitude higher gain for same length and/or same DC power supplied), and (ii) substantial gain-bandwidth product improvement for SIP schemes as well as (iii) significantly higher power efficiency (especially for high power gain values) compared to a single mode Pierce TWT modeled with one transmission line.

The rest of the paper is organized as follows: in Section II we present the background and the main idea to be developed in the paper. In Section III we summarize the theory of waveguides with SIP in the dispersion diagram and introduce the three, coupled transmission line (CTLs) model that is capable of exhibiting SIP [see Fig. 1(a) and (b)]. We also discuss engineering the SIP to have slightly positive or slightly negative slope (i.e., group velocity). We then consider finite-length SWS and inspect transfer functions, effects of loss and loading. In Section IV, we investigate the unique interaction scheme of such SWS with SIP with the electron beam where we study convective instability regime (i.e., amplification) and we briefly discuss instabilities associated with oscillator. We finally discuss the advantages of the proposed regime of amplification focusing on the three properties (i)-(iii) aforementioned. Throughout this paper, we assume linear operation of the TWT and assume that the time-dependence is implicit in the form of $e^{j\omega t}$ and hence suppressed.

## II. BACKGROUND AND PROBLEM STATEMENT

We investigate the simultaneous synchronization of the charge wave modulating the electron beam with *three* Floquet-Bloch eigenmodes with positive phase velocity along the $+z$-direction of a periodic SWS near the SIP. The simultaneous beam-synchronization of three electromagnetic modes affects amplification of the electromagnetic waves in an SWS. As discussed in [27] and in Sections III and IV, the SIP in the cold periodic, lossless, structure has a dispersion relation that, in the neighborhood of $\omega_{\mathrm{SIP}}$, is approximated by a third order power as

$$(\omega - \omega_{\mathrm{SIP}}) \approx h(k - k_{\mathrm{SIP}})^3 \qquad (1)$$

where $\omega_{\mathrm{SIP}}$ is the angular frequency at which three eigenmodes coalesce, $h$ is a constant that depends on the chosen SWS parameters, $k$ is the Floquet-Bloch wavenumber in the periodic and lossless SWS and $k_{\mathrm{SIP}}$ is the same at the SIP. As shown in the dispersion diagram of the fundamental Brillouin zone in Fig. 1(c), two different SIPs are possible at $\omega_{\mathrm{SIP}}$: (i) an SIP in the region $0 < k_{\mathrm{SIP}}d < \pi$ that exhibits positive group velocity for frequencies higher and lower than the SIP frequency and (ii) another SIP in the region $\pi < -k_{\mathrm{SIP}}d + 2\pi < 2\pi$ that is associated to negative group velocity for frequencies higher or lower than the SIP frequency. In the rest of the paper we are assuming that the electron beam is synchronized with electromagnetic modes with wavenumber in the range $0 < k_{SIP}d < \pi$ as shown in Fig. 1(c). Therefore the electron beam interacts with eigenmodes with SIP associated to the positive-slope branch of the dispersion diagram for frequencies in the vicinity of $\omega_{SIP}$. Indeed, *near* the SIP angular frequency $\omega_{SIP}$ there are three eigenmodes with distinct Floquet-Bloch wavenumbers $k$ provided by (1) where $h$ is positive. Two eigenmodes have complex conjugate wavenumbers and thus represent evanescent waves while the third one has a purely real Floquet-Bloch wavenumber and thus represents a propagating wave in the lossless SWS (the one shown in the upper branch of Fig. 1(c)). These three eigenmodes are independent in the sense that they have independent eigenvectors, i.e., different field distributions in the unit cell of the periodic structure for $\omega \neq \omega_{SIP}$. Note that the group velocity is always positive near $\omega_{SIP}$, contrarily to what happens in SWSs with DBE [31]–[33] where near the DBE four eigenwaves with both positive and negative group velocities are present. Exactly at $\omega = \omega_{SIP}$ the three eigenmodes associated to the SIP coalesce in their wavenumbers and eigenvectors; and the system matrix description of the periodic structure becomes defective [27], [40]. To provide a complete basis for the field propagating at the SIP; generalized eigenvectors shall be considered and thus such fields grow linearly and quadratically along the *z*-direction [34], [40] because of the special SIP dispersion.

To realize an SIP in a cold reciprocal SWS, three waves must be periodically-coupled. Therefore, the minimum realization of such phenomenon is achieved using three CTLs as in Fig. 1(a) that are able to support three eigenmodes in each direction, i.e., a total of six eigenmodes at each frequency if we account for both the $k$ and the $-k$ solutions. In contrast to the proposition in [27] that showed that one dimensional magnetic photonic crystals (MPCs) develop SIP and DBE characteristics, provided that nonreciprocal magnetic materials (ferrites for instance) are utilized. As such, in MPCs the dispersion diagram is asymmetric due to non-reciprocity. In [41] multi-ladder lumped circuit model is shown to emulate a $\omega - k$ dispersion diagram with a stationary inflection point.

We utilize here the three CTLs model in Fig. 1(b) to mimic the dispersion of an SWS with SIP. The series capacitance models cutoff frequencies in metallic waveguides (see Ch. 8 in [42]). The *generalized* Pierce theory developed in [22], [31], [32] allows the electron beam to interact with multiple waves in an SWS. The electron beam effect on the CTL is described using the equivalent, distributed, dependent current generators as seen in Fig. 1(b). See Refs. [31]–[33],[38] for more details on the generalized Pierce theory of multimodal periodic SWS interaction with an electron beam, including a Lagrangian formalism in [37]. We employ such general framework to investigate a specific SWS that develops an SIP. An example of the dispersion relation around the SIP is shown in Fig. 1(c) for a normalized case of the three CTLs shown in Fig. 1(a) where, as customary, only the real branches (i.e., with $k$ purely real, assuming absence of losses) of the $\omega - k$ diagram is plotted.

The plot also shows the so called "beam line" given by the charge wave dispersion relation $(u_0 k - \omega)^2 = 0$, where $u_0$ is the average speed of the electrons in the electron beam [3], [44]. The three-eigenmode synchronous interaction occurs when the beam line is chosen to intersect the SIP as seen graphically in Fig. 1(c). In other words, the proposed regime of operation mandates a *three-eigenmode synchronization* with the electron beam in which the phase velocity of the eigenmodes in the SWS equals the velocity $u_0$ of electrons, i.e., in the proposed regime the TWT design formula is

$$u_0 = \omega_{SIP} / k_{SIP}. \qquad (2)$$

In the following sections, we first investigate the characteristics of the cold structure exhibiting the SIP in terms of transmission and bandwidth; which are essential for achieving wideband amplification. Then, we show the amplification characteristic of the three-mode SWS synchronous regime with an electron beam. We explore mainly two cases of SWS interacting with electron beams: SWSs with SIP, and also SWSs with slightly positive dispersion slope (i.e. positive group velocity) at the frequency where modes tend to coalesce.

### III. COLD SWS FEATURING SIP

#### A. Three CTL Model and Cold SWS Characteristics

The proposed multimode periodically-coupled SWS featuring SIP is modeled by three CTLs as schematically shown in Fig. 1(a). In Fig. 1(a), each unit cell of the periodic three CTLs is composed of two successive CTL segments *A* and *B* and has the total length of *d*. In addition, each segment is described using TL distributed parameters including coupling capacitances and inductances per unit length in Fig. 1(b), whose values are reported in Appendix B.

In order to show the SIP in periodic structures and the slow-wave resonance behavior of the cold periodic CTLs with finite length ($L=Nd$), we use a transfer matrix procedure already explained in [22], [31], [32], [45]. In particular, we use a six-dimensional state vector $\boldsymbol{\Psi}(z) = \begin{bmatrix} \mathbf{V}^T(z) & \mathbf{I}^T(z) \end{bmatrix}^T$, where $\mathbf{V}^T(z) = \begin{bmatrix} V_1(z) & V_2(z) & V_3(z) \end{bmatrix}$, and $\mathbf{I}^T(z) = \begin{bmatrix} I_1(z) & I_2(z) & I_3(z) \end{bmatrix}$, that describes the evolution of the RF field amplitudes in the SWS. The evolution of this state vector from a coordinate $z_1$ to $z_2$ is described by $\boldsymbol{\Psi}(z_2) = \underline{\mathbf{T}}(z_2, z_1)\boldsymbol{\Psi}(z_1)$, where $\underline{\mathbf{T}}$ is a $6 \times 6$ transfer matrix [31], [32]. The six eigenmodes supported by three CTLs are provided by the eigenvalue problem

$$\left[\underline{\mathbf{T}}(z+d, z) - e^{-jkd}\underline{\mathbf{1}}\right]\boldsymbol{\Psi}(z) = 0, \qquad (3)$$

where $\underline{\mathbf{1}}$ is the $6 \times 6$ identity matrix. The six Floquet-Bloch wavenumbers are obtained by solving the dispersion equation

$$D_{SWS}(\omega, k) = \det\left[\underline{\mathbf{T}}(z+d, z) - e^{-jkd}\underline{\mathbf{1}}\right] = 0 \qquad (4)$$

discussed in the following sections and in Appendix A. The $\omega, k$-pairs solutions of (4) provide the dispersion diagram in Fig. 1(c), where, as customary, only the pairs with both real $\omega$ and wavenumber $k$ are plotted for simplicity. Note that (3) provides the six associated eigenvectors at any angular frequency different from $\omega_{SIP}$. However exactly at the SIP frequency there is only an independent eigenvector associated to $k_{SIP}$ and one to $-k_{SIP}$, and two generalized eigenvectors associated to each wavenumber [34], [40]. Generalized eigenvectors provide field solutions that grow linearly and quadratically along the *z*-direction [34], [40].

We find that, as an illustrative example, the periodic three CTLs model in Fig. 1(a), is able to support an SIP at $f_{SIP} = \omega_{SIP}/(2\pi) = 1.16$ GHz, when using the CTLs parameters provided in Appendix B. The coupling among the three CTLs are tuned such that its cold, lossless, dispersion relation features an SIP in the upper branch, as shown in Fig. 1(c) where only the modes with purely real $k$ (Floquet-Bloch wavenumber) are plotted.

Notice that the cold SWS is a reciprocal periodic structure, therefore if $k$ is a Floquet-Bloch wavenumber solution of the dispersion equation (4), $-k$ is also a solution of (4). Thus, the SIP also exists when the three modes with $-k$ coalesce at $-k_{SIP}$, still at $\omega = \omega_{SIP}$. Furthermore, because of periodicity, if $k$ is a solution of the dispersion equation, then also $k + m2\pi/d$ is a solution, where $m = 0, \pm 1, \pm 2, ...,$ is the Floquet harmonic index and $d$ is the period of the SWS. Therefore, if SIP exists at $\pm k_{SIP}$, an SIP is visible in the dispersion diagram at all the Floquet harmonics whose wavenumber is given by $\pm k_{SIP} + 2m\pi/d$. It is important to point out that with SWSs that support three eigenwaves in each direction the SIP occurs *inside* the Brillouin zone of the periodic structure; it does not occur at the Brillouin zone boundaries ($k = 0$ and $k = 2\pi/d$) nor at its center ($k = \pi/d$) hence it is not a band edge condition. The dispersion diagram in Fig. 1(c) provides the Floquet-Bloch wavenumber in the Brillouin zone defined here from $k = 0$ to $k = 2\pi/d$, keeping in mind that the dispersion is symmetric around any $k = m\pi/d$ which is a result of reciprocity of the cold SWS. Note that both the SIP frequency and the flatness of the dispersion curve at $\omega = \omega_{SIP}$ can be arbitrarily changed by tuning the coupling parameters between the three CTLs.

To further explore the properties of the CTLs with third order modal degeneracy in Fig. 1(a), we consider a finite-length cold CTLs structure consists of *N* unit cells as in Fig. 2(a). The cold voltage transfer function

$$TF_1 = V_{out1}/V_{in1}, \quad (5)$$

is obtained from the transfer matrix formalism. Here $V_{out1}$ is the voltage at the first TL output termination impedance $Z_{L1}$ and $V_{in1}$ is the input voltage of the first TL as shown in Fig. 2(a). Here we only excite the first TL as in Fig. 2(a), and we first assume that $Z_{s1} = Z_{L1} = 75\,\Omega$, while all the other input and output terminals are short circuited. The magnitude of the voltage transfer function $TF_1$ is plotted versus frequency for different number of unit cells (*N*) in Fig. 2(b), where results are shown for *N*=8 in dashed red, *N*=16 in solid blue, and *N*=24 in dotted black. It is observed that there are multiple resonance peaks near the SIP angular frequency $\omega_{SIP}$, and the main resonance denoted by $\omega_r$, is the sharpest and the closest to $\omega_{SIP}$, featuring higher quality- (*Q*-) factor than the others. As the number of unit cells (*N*) increases more and more resonance peaks are observed and the main one becomes even sharper.

### B. Engineering of the Dispersion Diagram with SIP

The transmission peaks in Fig. 2(b) are associated with the Fabry-Perot resonances of the finite cold periodic SWS. The sharper the resonance, the larger the group delay associated with it. This leads to high field enhancement and high *Q*-factors as the length of the SWS increases; which may also induce

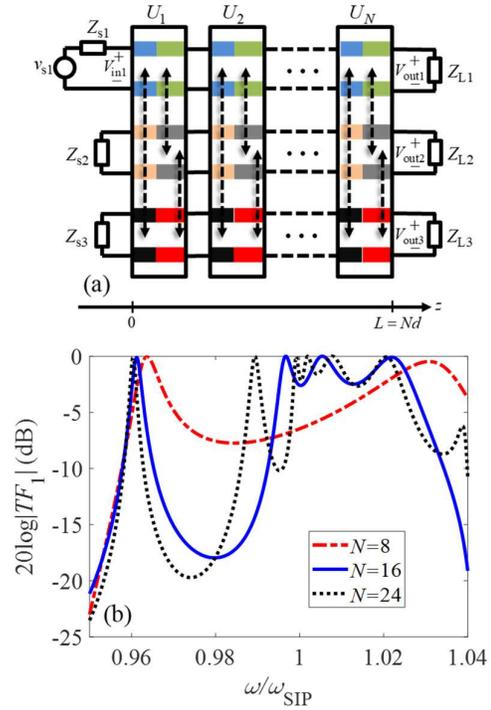

Fig. 2. (a) Finite length of a three CTLs periodic structure, with total length $L = Nd$ including loading and excitation, where *N* is the number of unit cells shown in Fig. (1). (b) Magnitude of the voltage transfer function $TF_1 = V_{out1}/V_{in1}$ for different number of unit cells (*N* = 8, 16, 24) with $Z_{s1} = Z_{L1} = 75\,\Omega$ and short circuit at all the other ports.

oscillations.

Depending on the target application, either to reduce the starting oscillation beam current (i.e., threshold current) for oscillators or to improve the gain and bandwidth for amplifiers, in this section we show how to synthesize and control the group velocity near the SIP frequency. By tuning the coupling amount between the TLs we can engineer the dispersion diagram to have small positive group velocities $(\partial\omega/\partial k > 0)$ or small negative group velocities $(\partial\omega/\partial k < 0)$ around the SIP frequency instead of the ideal case with zero group velocity (see Appendix A). The dispersion diagrams of the three cases near the SIP frequency are shown in Fig. 3(a) where we obtain a positive slope (dotted green curve) and negative slope (dashed red curve) group velocities around the SIP by having small changes in the coupling capacitors and inductors as provided in Appendix B. The magnitude of the voltage transfer function defined in (3) for these three cases is calculated and plotted in Fig. 3(b) for the case of having $N$=16 unit cells and $Z_{s1} = Z_{L1} = 75\,\Omega$ while the other TL terminals are short circuited.

By increasing the slope of the dispersion curve around the SIP frequency to achieve a positive group velocity, the resonance transmission peaks near the SIP frequency become wider compared to those for the ideal SIP having zero group

*C. Effects of Loss and TL Terminating Loads*

The $Q$-factor of a cavity made by a finite length SWS depends also on the terminating loads of the CTLs. A proper loading choice is of critical importance for the design of oscillators and amplifiers. Therefore, in this section, we examine different loading scenarios for the finite length CTLs with SIP. The magnitude of the voltage transfer function for the ideal SIP scheme is shown in Fig. 4(a) for three different loading scenarios. In the first loading scenario (blue solid curve), we terminate both input and output of the first TL with a resistive load ($Z_{s1} = Z_{L1} = 75\,\Omega$) while all the other TL terminals are short circuited. The second loading scenario [red dotted curve in Fig. 4(a)] is the case where the input terminal of the first TL and velocity. This is beneficial for achieving higher bandwidth in amplifier applications while retaining gain enhancement, and we will elaborate further on that in Section IV. On the other hand, by changing the coupling parameters we also obtain a negative slope in the dispersion diagram curve around the SIP frequency which results in sharper resonance peaks and consecutively higher $Q$-factors, as shown in Fig. 3(b). Therefore, such CTLs with locally negative slopes would be beneficial for oscillator applications which would be further discussed in Section IV, though the main focus of this paper is to discuss amplification properties of SWSs with SIP.

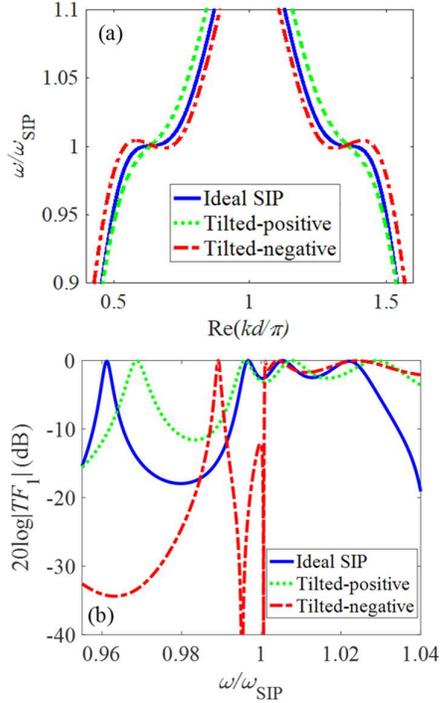

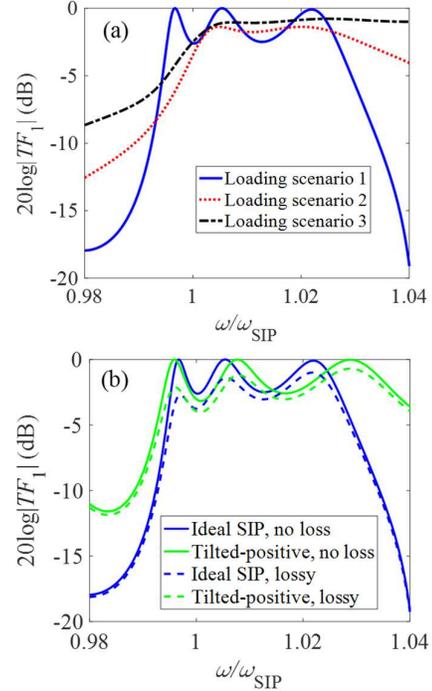

Fig.3. (a) Engineering the slope of the upper branch of the dispersion diagram in Fig. 1 to have positive and negative slopes around the SIP compared to the ideal SIP case of zero group velocity (in blue). (b) Magnitude of the voltage transfer function $TF_1 = V_{out1}/V_{in1}$ in dB for three different scenarios of ideal, tilted-positive and tilted negative dispersion relation around the SIP, for $N$=16 unit cells with $Z_{s1} = Z_{L1} = 75\,\Omega$ and short circuit at all the other ports.

Fig.4. (a) Comparison of three different loading scenarios for the finite three CTL structure. (b) Magnitude of the voltage transfer function of the two schemes of ideal SIP and tilted-positive for two cases of no loss and lossy structures when we have $N$=16 unit cells with $Z_{s1} = Z_{L1} = 75\,\Omega$ and short circuit at other ports as loading.

all the three output TL terminals are terminated by an equal resistive load ($Z_{s1} = Z_{L1} = Z_{L2} = Z_{L3} = 75\,\Omega$) while the input terminal of the two other TLs are short circuited. The third loading scenario (black dashed curve) is similar to the second one, i.e., $Z_{s1} = Z_{L1} = Z_{L2} = Z_{L3} = 75\,\Omega$ except that now the *input* terminals of the second and third TLs are open. It is observed that by terminating all the three output terminals with a load, the amplitude of the transfer function [defined in (5)] drops compared to the case where two output terminals are short circuited. However, the bandwidth increases dramatically specially for the third loading scenario where the second and third TLs' input terminals are left as open circuit. Therefore for amplifier applications since we are interested in having higher bandwidth, we use the third loading scenario here after. Note that the value of the resistive load would also affect the bandwidth. Here, we use the resistance loads of $75\,\Omega$ which seems providing the widest bandwidth. In the next section we will focus on this specific choice, though a deeper investigation of loading schemes in multiple CTLs could be beneficial to further maximize the bandwidth.

Next, we investigate the impact of the SWS losses on the transfer function. The presence of loss can impact the quality factor of the SWS structure and therefore its performance especially where we have high sensitivity to defects near the SIP. In Fig. 4(b) we plot the magnitude of the voltage transfer function for the two SWS with ideal and tilted-positive SIP with and without considering loss effects. Here we have assumed the loss as a distributed series resistance of $R_{\mathrm{loss}} = 5\,\Omega/\mathrm{m}$ in each segment of the periodic three CTLs, composed of $N=16$ unit cells. The results in Fig. 4(b) show how the magnitude of the transfer function as well as the $Q$-factor decrease as compared to that in the absence of losses as expected, for both tilted and ideal SIP, when losses are accounted for.

## IV. HOT SWS FEATURING SIP

We investigate here amplification in TWTs modeled as three CTLs interacting with an electron beam. We use the generalized Pierce theory developed in [31], [32] for multi transmission lines interaction with an electron beam. Accordingly, the interaction of the SWS modes with an electron beam is taken into account by modeling the beam as a continuous flow of charges, i.e., by charge waves as in [3], [22]. Such charge wave describes the bunching and debunching of the beam's electrons, which causes energy exchange between the beam and the modes in the SWS. The electron beam has an average (d.c.) current $-I_0$ ($I_0$ is a real, positive number) along the $+z$ direction and a d.c. equivalent beam voltage $V_0$ that is related to the average electron velocity $u_0$ by $V_0 = u_0^2/(2\eta)$, where $\eta = e/m = 1.758 \times 10^{11}$ C/kg is the electron charge-to-mass ratio, where $-e$ and $m$ refer to the charge and mass of the electron, respectively. The electromagnetic fields in the waveguide induce a perturbation (modulation or disturbance) on the charge wave [3] described by a modulation of the charge wave current, $I_b$, and modulation of beam velocity, $u_b$, and equivalent kinetic voltage modulation $V_b = u_0 u_b/\eta$ [22], [31], oscillating at the same frequency as the electromagnetic fields in the SWS. Therefore the total beam current along the positive $z$ direction is $-I_0 + I_b$ with $|I_b| \ll I_0$ and the total equivalent beam voltage $V_0 + V_b$ with $|V_b| \ll V_0$, for small-signal considerations. We conveniently define a space-varying *eight-dimensional state vector* composed of all the field quantities that vary along the $z$-direction, which are the transmission line voltage and current vectors already discussed in Sec. III, as well as the charge wave modulation terms $I_b$ and $V_b$. In summary, the system state vector made of eight components is

$$\Psi(z) = \begin{bmatrix} \mathbf{V}^T(z) & \mathbf{I}^T(z) & V_b(z) & I_b(z) \end{bmatrix}^T \quad (6)$$

and it describes the evolution of time-harmonic electromagnetic waves and beam charge wave along the $z$-direction using our multi TLs approach [22], [31], [32].

In order to investigate the oscillation and amplification behavior of the proposed three CTLs with SIP, in Sec. IV-*B*, we consider a finite length SWS composed of $N$ unit cells as in Fig. 5 with appropriate source and loads. The power gain transfer function is defined as

$$G_{\mathrm{P1}} = P_{\mathrm{L},1}/P_{\mathrm{smax}} \quad (7)$$

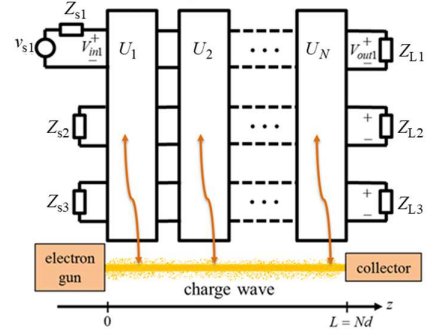

Fig. 5. Conceptual example of a three CTL structure interacting with an electron beam with the generalized Pierce model of the three CTLs, with unit cells as shown in Fig. 1(b) including the per unit length parameters. The blocks $U_n$ represent unit cells made of the two segments $A$ and $B$ in Figs. 1(a) and 2, with parameters given in Appendix B.

where $P_{\mathrm{smax}} = |v_{s1}|^2 / [8\mathrm{Re}(Z_{s1})]$ is the maximum available power at the input terminals of the first TL and $P_{\mathrm{L},1}$ is the power delivered to the load $Z_{\mathrm{L1}}$ of the first TL, since it is the port where most of the power is delivered as it will be clarified later on.

### A. Dispersion Relation of the Combined SWS-Electron Beam Interactive System

In this section, we analyze the dispersion relation of modes in a SWS, designed to have a SIP, interacting with an electron beam. We investigate the effect of the electron beam current on the dispersion diagram assuming the real frequency and a wavenumber $k$ that can have complex values. Modes in the cold SWS are degenerate and satisfy the synchronous condition (2). The addition of gain perturbs the degeneracy condition though the degeneracy structure is somewhat preserved when gain levels per unit length are not so high. Here and in Appendix A we show the complex $k$ dispersion diagram varying the electron beam current and observe how much the SIP condition is perturbed from the "cold" case.

The eight complex $k$ modal solutions at any real angular frequency $\omega$ are found by solving for the eigenvalues of a system as in (3), where now $\Psi(z)$ is the 8-dimensional state vector given in (6) and $\mathbf{T}(z+d,z)$ is a 8×8 system matrix assocviated to the periodic CTLs interacting with the electron beam as discussed in [22], [31], [32]. In general, one can write the dispersion equation of the coupled SWS-electron beam system, simply as

$$D_{\text{SWS}}(\omega,k) P_{\text{beam}}(\omega,k) = C(\omega,k,I_0) \quad (8)$$

in which $D_{\text{SWS}}(\omega,k) = 0$ is the dispersion relation of the cold SWS given in (4), and $P_{\text{beam}}(\omega,k) = 0$ is the dispersion relation of the cold beam [44], namely

$$P_{\text{beam}}(\omega,k) = (u_0 k - \omega)^2 - \Omega_{\text{P}}^2 = 0 \quad (9)$$

where $\Omega_{\text{P}}$ is the plasma angular frequency, which includes space charge effects [3], [9]. Note that (9) provides two solutions $k = \omega/u_0 \pm \Omega_{\text{P}}$ for the electron beam-alone modal solutions of the modulated charge wave; these solutions are positive and do not satisfy reciprocity, i.e., $k$ and $-k$ are not both solutions of (9).

In the CTL-beam system, the function $C(\omega,k,I_0)$ on the right hand side of (8) represents phenomenologically the coupling between the electromagnetic modes of the cold SWS and those that would be supported by the electron beam alone. In simple cases such as in the original Pierce model, $C(\omega,k,I_0)$ can be written in an exact analytical form, (see for instance [3], [44] or [46]). The function $C(\omega,k,I_0)$ also might be understood as a perturbation term of the SWS system due to the electron beam.

Note that one can always cast the dispersion of the interactive systems in the form given by (8). However, for the investigated cases reported here, where periodicity is involved and the SWS is made of three TLs, it is extremely tedious to find analytical expressions for $D_{\text{SWS}}(\omega,k)$ or $C(\omega,k)$ since they involve the determinant of the 8×8 system matrix. On the other hand, it is useful to investigate the interaction in the vicinity of the SIP where the dispersion of the cold SWS is approximated by the SIP dispersion in (1) as shown in Appendix A.

Since synchronization between the electron beam and the CTLs occurs around the SIP frequency [see (2)], where three SWS eigenwaves coalesce, a strong interaction between the electron beam and the SWS modes is expected. As mentioned earlier we assume the beam interacting with the three degenerate eigenwaves associated to the SIP on the left side of the dispersion relation in Figs. 1(c) and 6 to exploit positive group velocity for high power amplification. To show what happens to the dispersion diagram of the hot SWS (in the presence of the electron beam) we plot in Fig. 6 the real part of $k$ for both cold (blue curves) and hot structures (red curves) in a wide frequency range around the SIP frequency. Blue curves represent the dispersion diagram of the "cold" SWS and are covered by the red curves (blue curves are coincident with those in Fig. 1), except for a region close to the interaction point ($k_{\text{SIP}}, \omega_{\text{SIP}}$). There, one can see the deformation and splitting of the modal curves around the point where synchronization with the electron beam happens. However, away from the interaction point ($k_{\text{SIP}}, \omega_{\text{SIP}}$) the modes (red curves) asymptotically approach those of the cold structure (blue curves) because waves in the CTLs are no longer synchronized with the electron beam. For beam currents larger than that considered in Fig. 6 the deformation of the dispersion diagram of the hot structure from that of the cold one extends significantly further from the SIP point. In any case the diagram is not symmetric with respect to $k = \pi/d$. A comprehensive investigation of the impact of the beam average current $I_0$ on the dispersion diagram for the hot structure is carried out in Appendix A.

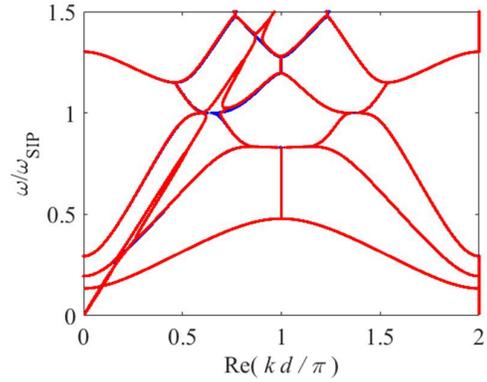

Fig. 6. Dispersion diagrams showing the *real* part of the complex Floquet-Bloch wavenumber $k$ versus angular frequency, for both the "hot" and "cold" systems. Dispersion curves for the "hot" SWS-electron beam interactive system (shown in solid red lines) are calculated using an electron beam current of $I_0 = 0.5$ A. Dispersion curves for the "cold" lossless SWS are shown with solid blue lines that are under the red curves (hence not visible) except for a small region near the interaction point ($k_{\text{SIP}}, \omega_{\text{SIP}}$) given in (2). Zooms and more detailed plots of the dispersion diagram for various beam currents are reported in Appendix A.

### B. Study of Instabilities at The SIP

In an infinitely-long SWS interacting with an electron beam, the oscillation condition and the starting oscillation current can be calculated from the dispersion diagram of the structure as explained in Section IV. In an infinitely-long structure with the e-beam as a source of linear gain, the electromagnetic waves exhibit either unbounded growing oscillations in time at every fixed point in space, or growing oscillations only in space, i.e., progressing along *z*; those instabilities are commonly referred to as *absolute and convective instabilities, respectively* [47], [48]. Criteria for these two distinct instabilities in an infinitely-periodic interactive system can be drawn from investigating the dispersion diagram thereof following the Briggs-Bers condition [49], [50].

Amplification at a given real frequency is due to convective instability, i.e., one or more eigenwaves of the infinitely-long SWS coupled to the electron beam, whose propagation is described by $\exp(-jkz)$, is exponentially growing while they progress along *z*, hence their complex wavenumber should be such that $\text{Im}(k) > 0$. Even though a system made of an *infinitely-long* SWS interacting with an electron beam does not exhibits absolute instability, there is another reason of why a system of *finite length* with internal amplification (i.e., convective instability) could start to oscillate. Growing oscillations could occur when amplified waves provide constructive interference with positive feedback after reflections at the two ends of the SWS with gain. Condition for such occurrence of oscillations in a *finite length* structure is examined by tracing the location of the poles of the transfer function (i.e., (5)) in the complex angular frequency $\omega$-plane as a function of the electron beam current $I_0$. In the following, the synchronous condition for RF amplification is assumed by choosing the velocity $u_0$ of the electrons to match the phase velocity of the three degenerate forward eigenmodes in the +*z*-direction according to (2).

For a stable (i.e., amplifying) system, all the system poles have positive imaginary parts in the complex $\omega$-plane.

The system starts to oscillate whenever one of the $\omega$–poles moves to the part of the $\omega$-plane with negative imaginary part. Here, for simplicity we consider the voltage transfer function defined in (5) relative to a system made of an electron beam interactive with three lossless CTLs with load terminations of loading scenarios 1 and 3 introduced in previous section. Although the *TF* in (5) exhibits many poles in the $\omega$-plane, the pole closest to the SIP frequency (represented by $\omega_p = \omega_{pr} + j\omega_{pi}$) is of special interest. This is the $\omega$–pole that transitions from the positive imaginary part of the $\omega$–plane to the negative imaginary part (unstable region) with the increase of the electron beam current $I_0$, much faster than the other poles of the transfer function. Hence, it is the pole that is responsible for starting oscillations. For an amplifier regime, all poles should be in the stable region. Fig. 7 shows the trajectory of the imaginary part of such a pole as a function of beam current $I_0$, for four different scenarios, all with a SWS made of *N*=16 unit cells. We are considering two loading scenarios here: (loading scenario 1) terminating both input and output of the first TL with a resistive load of $75\Omega$ while all the other four TL terminals are short circuited, and loading scenario 3) terminating the first TL input and all the three TL output terminals by an equal resistive load of $75\Omega$ while the input terminal of the two other TLs are left open. The starting oscillation beam current ($I_{\text{th}}$) is the beam current value for which the pole enters the unstable region $(\text{Im}[\omega_p] < 0)$ from the stable region $(\text{Im}[\omega_p] > 0)$. In other words, the pole crosses the real $\omega$ axis if a strong enough electron beam current is used in the SWS device, and after that the system will start oscillations.

We observe from Fig. 7 that the SWS with titled SIP regime with *negative group velocity* (red solid curve) has the closest pole to the real $\omega$ axis which moves very quickly into the oscillation region by slightly increasing the beam current. Consequently, it has the lowest starting oscillation electron beam current and is best suited for oscillator designs. On the other hand, the SWS based on the regime with titled SIP with *positive group velocity* (dashed green curve) has the largest starting oscillation beam current compared to the other regimes. Hence the SWS with titled positive slope regime is best suited (compared to the other cases in Fig. 7) for the high power amplifier design. In the next subsection, we specifically examine the potential applications of SWS with ideal and titled SIP with positive group velocity for conceiving amplifier regimes, i.e., regimes based on convective instability. We observe from Fig. 7 that the starting oscillation beam currents for ideal SIP and tilted-positive SIP with the loading scenario 3 are $I_{\text{th}} = 0.56$ A and $I_{\text{th}} = 0.94$ A, respectively, and in all of our results hereafter we consider electron beam currents much smaller than these starting currents to avoid oscillations.

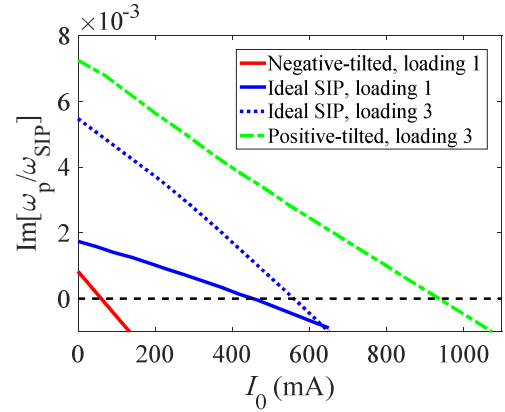

Fig. 7. Imaginary part of the pole nearest to SIP plotted versus beam current $I_0$ for four different scenarios: (i) tilted-negative regime (oscillator application) assuming loading scenario 1; (ii) ideal SIP regime assuming loading scenario 1; (iii) ideal SIP regime assuming loading scenario 3; (iv) tilted-positive regime (amplifier application) assuming loading scenario 3. The system is unstable when a pole has $\text{Im}[\omega] < 0$. The "tilted-positive" case is the most stable.

### C. Amplification

The power gain defined in (7) is plotted in Fig. 8(a) versus normalized angular frequency near the SIP frequency for both amplifiers based on the ideal SIP regime and the tilted-positive one. The beam current value for each case is chosen such that the maximum power gains for both cases are equal to either 10 dB or 25 dB. In addition, the electron beam current values are chosen such that the structure operates in the stable region, so in each respective case we have $I_0 < I_{\text{th}}$.

In our simulations for SIP amplifiers we assume loads of $75 \, \Omega$ at every output port while the two lower input terminals are left open circuited (loading scenario 3), since this configuration provides a large enough bandwidth to prove benefits, though optimization of the regime of operation would lead to even better results. It is observed from Fig. 8(a) that the amplifier based on tilted SIP regime with the positive group velocity requires a larger beam current as compared to the amplifier based on an ideal SIP to provide the same amount of the power gain. Although the amplifier with tilted SIP with positive group velocity features a lower gain, it provides a higher bandwidth as compared to an ideal SIP case.

Next, we investigate in Fig. 8(b) the distribution of output power $P_{L,i}$, with i = 1,2,3, on the output CTLs terminals. The power gain of each output is defined as $G_{P,i} = P_{L,i} / P_{smax}$. In addition, the total power gain, i.e., $G_P = \sum_i P_{L,i} / P_{smax}$, is also plotted for comparison. It is observed from Fig. 8(b) that most of the available power is delivered to the first TL's output terminal and very small portions of the available power is transferred to the other two TLs' output terminals. Though this

In Fig. 9(a), we plot the total power gain ($G_P$) relative to the amplification regimes with ideal SIP and tilted SIP with positive group velocity versus electron beam current $I_0$. To provide a better assessment on the performance of the SIP regimes of operation, we also plot the power gain in a TWT amplifier made of a single periodic TL having the same total length and the same characteristic impedance as the one in the TLs used for the SIP cases. The SWS made of single TL is assumed to be periodic to provide a fair comparison, since results are compared to those of amplifiers made of periodic SWS with SIP. The parameters of the single TL interacting with the electron beam are provided in Appendix B, and the results are based on a Pierce model for the single mode SWS made of one TL interacting with the electron beam, with calculations done using a transfer matrix formalism.

Note that for each amplification regime the range of the beam current is such that $I_0 < I_{th}$ so it operates in the stable

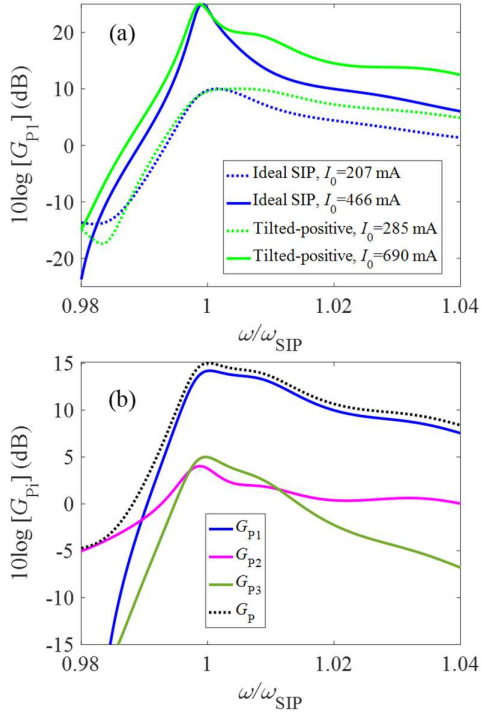

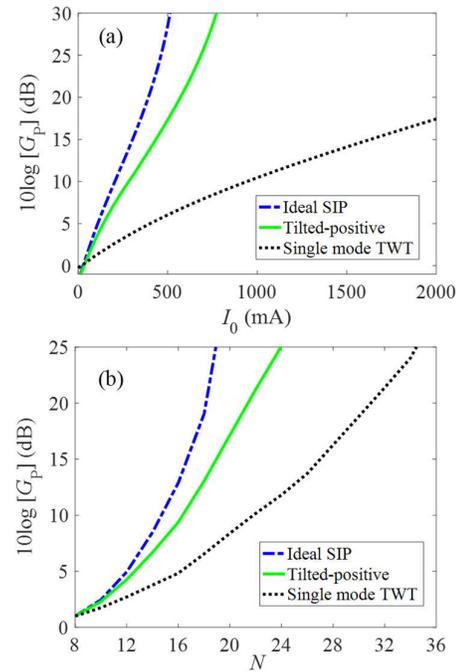

Fig. 8. (a) Comparison of the power gain $G_{P1} = P_{L,1} / P_{smax}$ for both the ideal and the tilted-positive SIP regimes interacting with an electron beam in which the value of beam current is selected such that the maximum power gain is set to be either 10 dB or 25 dB. (b) Power gain at each of the three output ports and the total output power gain $G_P$, for a regime of operation based on the SWS with tilted-positive group velocity. For both plots the gain is calculated for a SWS with *N*=16 unit cells with $Z_{Li} = 75\Omega$, *i*=1,2,3, $Z_{s1} = 75\Omega$ and open circuit for the two other input ports. We show only convective instability regimes of operation.

condition may vary depending on the adopted CTL configuration.

Fig. 9. (a) Power gain (in dB) of the proposed ideal SIP and tilted-positive regimes of TWT operation, compared with a standard regime of amplification based on a finite, periodic TL, modeling a single mode TWT, having the same characteristic impedance, length and loading plotted versus the beam current. Results are based on structures with *N*=16 unit cells, with loads of $Z_L = 75\Omega$ (loading scenario 3) and $R_{loss} = 5 \, \Omega/m$. (b) Total power gain versus the number of SWS unit cells, *N*. Comparison among amplification regimes based on SWS with the ideal SIP, the titled-positive dispersion, and a single TL model introduced in part (a). The values of the electron beam current is selected for each regime such that the same gain is achieved for the case of *N*=8 unit cells. The resulting electron beam current values for the ideal and tilted-positive SIP is 263 mA, in both cases, whereas for single TL regime the beam current is 385 mA. The threshold beam currents ($I_{th}$) for ideal, tilted-SIP, and 1TL are 560 mA, 880 mA and 4.7 A respectively.

also that the gain for the amplifier with tilted SIP with positive group velocity is still much larger than the single TL case.

The power gain of the proposed tilted-positive SIP amplifier is plotted in Fig. 9(b) versus the length of the periodic SWS with $N$ unit cells, and compared it to that of the single TL amplifier. This figure shows a huge advantage for the proposed amplifiers based on the ideal SIP and tilted-positive SIP regimes in terms of power gain as $N$ increases, when compared to the single TL.

As an important figure of merit for amplifiers, we plot in Fig. 10(a) the gain-bandwidth product ($G \times BW$) results for the ideal SIP and the tilted-positive amplification regimes as compared to that of the periodic single TL pierce model versus electron beam current. In this figure, the $G \times BW$ is defined as the product of the maximum power gain and the 3dB-gain bandwidth. The transmission lines parameters as well as loading for the three regimes of operation are provided in the Appendix B and also in the caption of Fig. 10. The results of Fig. 10(a) show that both the SIP amplification regimes outperform the conventional regime based on the single TL Pierce model, for electron beam currents larger than $I_0 = 300$ mA where the gain is high enough to be desirable for amplifier applications.

In addition, we plot in Fig. 10(b) the 3dB-gain bandwidth versus power gain for the three regimes of amplification discussed previously. Different values of gain are obtained by changing the electron beam current $I_0$. It should be noted that in order to have equal power gain for the three cases mentioned above, the values of the electron beam current are different for each case which results in different power added efficiency for each case as discussed next. The first and foremost observation of the Fig. 10(b) is that although higher bandwidths is expected for the single TL amplifier, for similar gain the tilted-positive SIP amplifier has slight higher bandwidth, within the 15 to 20 dB gain range. Though bandwidth of the new SIP amplification regimes is not higher than that of the single TL amplifier in these results (proposed here for the first time), if we look at the $G \times BW$ figure of merit in Fig. 10(a), it is clear that both SIP regimes of amplification provide higher values than the single TL regime. Furthermore, it must be noted that the amplifier based on the tilted-positive SIP regime, though it provides smaller gain than the ideal SIP regime, it provides higher bandwidth.

Finally, we examine another important parameter for amplifiers, the power added efficiency (PAE), which is defined (in percentage) as

$$\text{PAE} = \frac{P_{\text{out(RF)}} - P_{\text{in(RF)}}}{P_{\text{DC}}} \times 100, \qquad (9)$$

where $P_{\text{DC}}$ is the d.c. electron beam power (defined as $P_{\text{DC}} = V_0 I_0$), whereas $P_{\text{in(RF)}}$ and $P_{\text{out(RF)}}$ are the input and total output RF powers respectively. In all cases, we use the same $P_{\text{in(RF)}} = (1/2)\operatorname{Re}\{V_{\text{in1}} I_{\text{in1}}^*\}$, and $P_{\text{out(RF)}} = G_p P_{\text{in(RF)}}$. The total output power is on the three loads, though the power region. From Fig. 9(a) we observe that, for the same amount of electron beam current, the amplifier based on the ideal SIP regime provides the highest power gain as compared to the two other cases. However, the amplifier with ideal SIP has a very limited bandwidth as compared to the amplifier with tilted-positive SIP as it is going to be discussed in the following. Note on load 1 is dominant. We observe from Eq. (9) that when comparing the three regimes with equal gain, the larger the required beam current $I_0$ for a given gain value, the smaller the PAE. Hence, the SWS-beam regime that requires the smaller beam current, at equal gain, has the largest PAE. In Fig. 10(c) we have plotted the PAE for the three SWS regimes mentioned above versus the power gain (power gain is varied by changing the electron beam current $I_0$). The foremost observation is that amplifiers based on the SIP regimes of operation feature significantly higher PAE as compared to the amplifier based on the single TL regime of operation, in particular at large power gains.

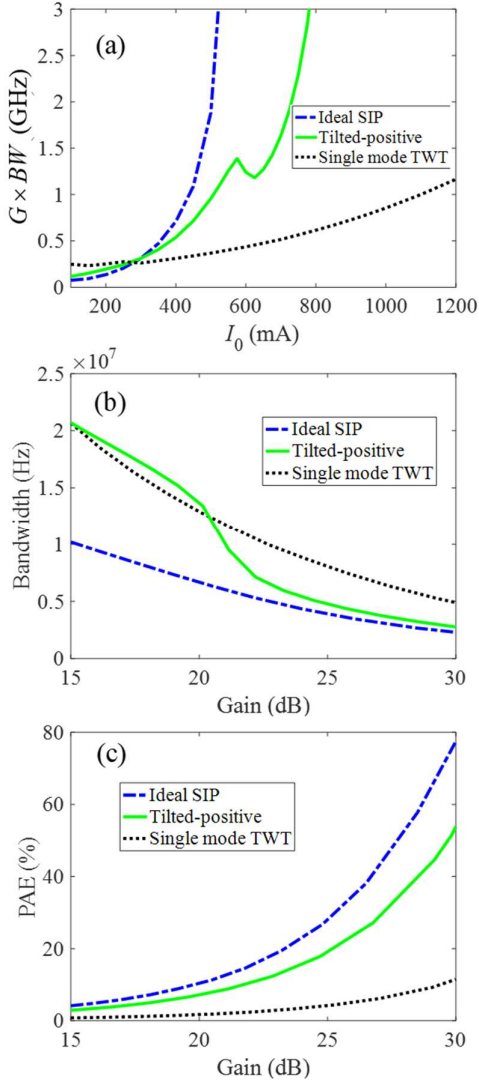

Fig. 10. (a) Comparison of the gain-bandwidth product ($G \times BW$) of TWT amplifiers based on SWS regimes based on ideal SIP and tilted-positive dispersion, versus electron beam current $I_0$. Results are also compared with a TWT made of a SWS with only one periodic TL (single mode TWT) having the same length and average characteristic impedance. (b) Bandwidth comparison for the three aforementioned regimes of operation versus power gain (in dB). (c) Power added efficiency (PAE) in percentage for the three regimes as a function of power gain (in dB). In (b) and (c), different beam currents are chosen for the three regimes such that all three cases have equal gains. In all cases we are assuming a SWS made of $N=16$ unit cells and the per unit length resistance of $R_{\text{loss}} = 5$ $\Omega/$m with $Z_{\text{Li}} = 75\,\Omega$, i=1,2,3, $Z_{\text{s1}} = 75\,\Omega$ and open circuit at the two other input ports for the two SIP regimes, and $Z_L = Z_s = 75\,\Omega$ for the single TL regime.

## V. CONCLUSION

We have reported a new regime of operation for SWSs with a third order degeneracy interacting with an electron beam with possible application in TWT amplifiers. The new regime of operation is based on a strong synchronous interaction between an electron beam and three Floquet-Bloch eigenmodes in a SWS having a stationary inflection point (SIP). Previously, in [31]–[33] we have analyzed a regime of operation where the electron beam was interacting with four degenerate eigenmodes at the DBE, which is a fourth order degeneracy at which eigenmodes have zero group velocity. Both positive and negative group velocities exist at frequencies near that of the DBE (similar to other band edge conditions, such as the RBE). The DBE was proposed to conceive low-threshold or high-efficiency oscillators. Instead, the SIP interaction regime for amplifiers investigated here does not exploit electromagnetic band gaps.

The proposed SIP-based regime of amplification utilizes only vanishing or slightly positive group velocity for frequencies either higher or lower than the SIP one leading to high power amplifiers. Note that the considered SIP associated with positive group velocities is the one shown in Figs. 1(c) and 6 occurring in the range $0 < kd < \pi$. Though not shown here one could conceive also a regime where the electron beam is synchronized with the SIP in the range $\pi < kd < 2\pi$, the range of wavenumbers associated with negative group velocity for backward wave oscillator designs, which are left for future investigation.

We have demonstrated that the new proposed amplification regime based on SIP provides a route for three improvements in TWT amplifiers: large gain, large gain-bandwidth product ($G \times BW$) as well as higher power efficiencies in TWTs compared to a Pierce-like TWT regime of operation (modeled as a single TL in this paper) having the same length and electron beam current.

Besides amplifiers based on SWSs with ideal SIP condition we have explored also one regime of operation where dispersion engineering leads to a modified SIP condition where group velocity is non-vanishing and always slightly positive. The important feature of the proposed scheme is that the slope of the dispersion relation can be engineered for enhancing the bandwidth. We have shown that the SIP with a slightly positive group velocity results in a higher bandwidth at the cost of gain/efficiency.

We have investigated how the positive group velocity of the three eigenmode degeneracy exploited in this paper makes these TWTs less prone to start oscillations compared to the regimes with vanishing group velocity (Fig. 6). Indeed the proposed SIP regime with vanishing or slightly positive group velocity is less prone to start oscillations than the previously studied regime based on the DBE [33] whose degeneracy involved also Floquet-Bloch modes with negative group velocity. We have also shown that we can engineer the SIP regime to have a slightly negative group velocity in the region $0 < kd < \pi$ and this provides significantly lower oscillation threshold than the SIP counterpart with positive group velocity, and may be beneficial for low threshold oscillator designs which can be a future research venue.

Despite the SWS with SIP dispersion is here based on an ideal periodic CTLs, realistic metallic waveguides with SIP can be devised analogously to what was previously demonstrated theoretically [35] and experimentally [51]: such circular metallic waveguides are able to support the DBE. The strategy to obtain realistic SWSs with SIP is to consider a periodic waveguide that supports three modes that are periodically coupled by modulating the shape in the longitudinal direction. Nevertheless, here we have shown for the first time some possible advantages of a new regime of operation for TWT amplifiers that could benefit microwave and millimeter wave generation. Precise engineering of such a multimode metallic waveguide with SIP is left to a future investigation now that we have demonstrated the possible advantages of TWT amplifiers based on SWSs exhibiting ideal SIP or SIP with positive tilt.

APPENDIX A: DISPERSION RELATION FOR INTERACTION OF SWS WITH ELECTRON BEAM

In this Appendix, a detailed investigation of the dispersion relation of the hot SWS structure (the combined SWS-electron beam system) is offered and we observe how the complex wavenumber $k$ is affected by the strength of the electron beam current $I_0$.

As mentioned in Section IV, the general dispersion equation of the coupled SWS-electron beam system is written as in Eq. (8) with the beam dispersion relation provided in Eq. (9). Therefore complex wavenumber solutions of Eq. (8) for real frequency provide the eight curves of the modal dispersion diagram in Fig. 6. In Eq. (8) the dispersion diagram of the combined structure is obtained by modifying the dispersion of the cold SWS. Therefore it is important to characterize the dispersion of the cold SWS for the ideal SIP and also for the titled cases described in this paper. In the following $D_{\text{SWS}}(\omega, k) = D_{\text{SIP}}(\omega, k)$ represents the case of ideal SIP, whereas $D_{\text{SWS}}(\omega, k) = D_{\text{SIP,titled}}(\omega, k)$ represents a SWS that instead of the ideal SIP expresses a slightly *tilted* version of it. Note that the dispersion diagram for a SWS that exhibits an ideal SIP, can be approximated as $D_{\text{SIP}}(\omega,k) \approx (\omega - \omega_{\text{SIP}}) - h(k - k_{\text{SIP}})^3 = 0$ in the vicinity of the SIP, which is coincident with Eq. (1). It is noteworthy to show that after some algebraic manipulations the dispersion relation for the *titled* SIP can be approximated as a perturbation of $D_{\text{SIP}}(\omega,k)$ as

$$D_{\text{SIP,titled}} \approx D_{\text{SIP}} - v_t k = \\ = (\omega - \omega_{\text{SIP}}) - h(k - k_{\text{SIP}})^3 - v_t k = 0, \quad (A1)$$

where $v_t$ is the minimum group velocity (in magnitude) of titled SIP cases, that occurs at the original SIP wavenumber $k = k_{\text{SIP}}$. In other words, the minimum group velocity of the titled-SIP mode occurs at $k = k_{\text{SIP}}$, and based on (A1) is given by

$$\left.\frac{d\omega}{dk}\right|_{k=k_{\text{SIP}}} = v_t \quad (A2)$$

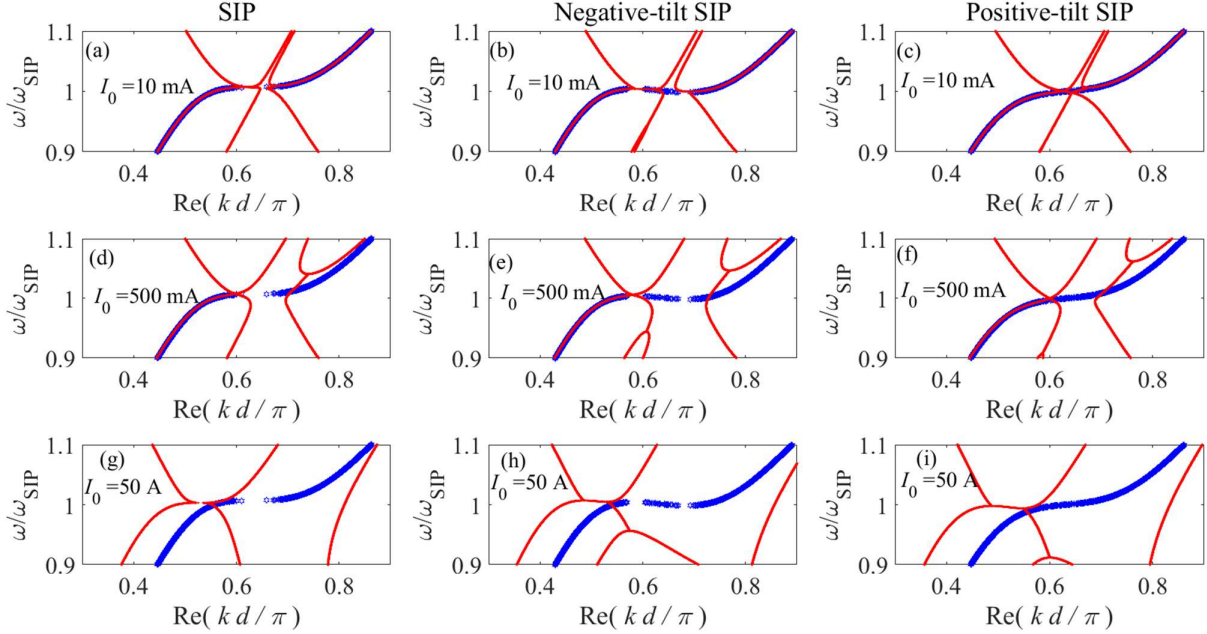

Fig. A1. Dispersion diagrams for the SWS-electron beam interactive "hot" system. In particular, we show the evolution of the *real* part of the complex Floquet-Bloch wavenumber *k* versus angular frequency for the three cases: ideal SIP, negative and positive tilted SIP, features for the cold SWS; considering different values of the beam current: for 10 mA in (a, b, c), for 500 mA in (d, e, f), and for 50 A in (g, h, i). The SWS structure is made of three CTLs. Solid red lines represent the dispersion of the interacting "hot" structure, whereas blue symbols represent the eigenmode with purely real wavenumber in the "cold" lossless SWS structure.

The formula in (A1) assumes that the tilt results in a small but non-vanishing group velocity at $\omega = \omega_{\text{SIP}}$, for which the SWS retains some of the degeneracy features within the limit of small perturbation. For $v_t > 0$ the dispersion has a positive slope and that is what we proposed in a TWT amplifier, while for $v_t < 0$ the dispersion has negative slope at the original SIP angular frequency $\omega_{\text{SIP}}$, and this regime could be used for low-threshold oscillators.

Next, we analyze the interaction between the SWS made of three TLs with SIP and the electron beam through the modal dispersion diagram calculated using the state vector defined in (6) and the associated 8×8 transfer matrix **T** for a unit cell as investigated thoroughly in [31]. Complex modes, i.e., modes with complex wavenumber *k*, are found by numerically solving

$\det\left[\mathbf{T}(z+d,z) - e^{-jkd}\mathbf{1}\right] = 0$ that is equivalent to (8). In Fig. A1 we plot the complex dispersion diagram (red curves) of such interactive structures for three different values of electron beam current $I_0$. Since we focus on amplification regimes we look for complex *k* solutions of (8) with real $\omega$. Therefore, Floquet harmonics that provide amplification along the positive *z* direction propagate as $\exp(-jkz)$ with $\text{Im}(k) > 0$. In our CTL interacting with the electron beam there could be more than one eigenwave that contributes to amplification.

We consider three regimes of operations: an SWS with an SIP, and two others with positive and negative-tilt in the dispersion diagram of the "cold" SWS structure. Modes in the cold SWS are denoted with blue symbols. A strong interaction with electron beam occurs due to synchronization between the CTLs and electron beam around the SIP point $(\omega_{\text{SIP}}, k_{\text{SIP}})$ which results in the deformation of the dispersion curves as a function of the electron beam intensity. While in Fig. 6 we show the diagram for a wide frequency range, in Fig. A1 for clarity we show what happens around $\omega_{\text{SIP}}$ by plotting the real part of $k$ in the normalized wavenumber range $0.35\pi < kd < 0.85\pi$, to show how modal curves deform and even split in that region. Notice how the SIP modes in the "cold" CTLs cases (curves with blue symbols) are deformed when the SWS interacts with the electron beam, and how this perturbation increases with increasing the beam current intensity. At each frequency, there are eight Floquet-Bloch modes solutions in the dispersion diagram for the "hot" SWS structure (i.e., the CTLs interacting with the beam), but only five complex ones are shown in Fig. A1 since the other three have negative Re($k$). Three curves are originated by the SWS (i.e., from $D_{\text{SWS}}(\omega,k)$ in (8)) and two are originated by the charge wave dispersion relation (i.e., $P_{\text{beam}}(\omega,k)$ in (8)), but they are perturbed and connected to each other because of the term $C(\omega,k,I_0)$ on the right side of (8).

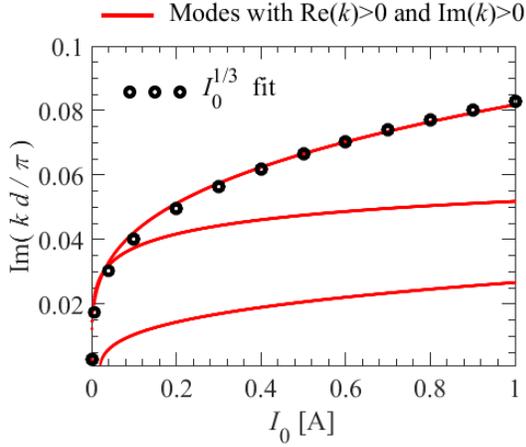

Fig. A2. Imaginary part of the Floquet-Bloch wavenumber, at fixed angular frequency, of the three amplifying modes having Re($k$) > 0 and Im($k$) > 0 of the "hot" SWS with SIP interaction with an electron beam varying as a function of the beam current. Note that there are five modes having Re($k$) > 0, but only three of them provide amplification since they have Im($k$) > 0. The chosen frequency of operation corresponds to that of the SIP frequency of the cold SWS. The imaginary part which is responsible for amplification for one of the modes grows with the beam current as $I_0^{1/3}$.

In some cases, only three distinct curves are visible, i.e., three values of Re($k$) > 0 at each frequency, which means in that case at least two curve-branches have two distinct values of Im($k$), for the same value of Re($k$).

For electron beam d.c. current $I_0 = 10$ mA we see that the SIP condition (solid curves) in Fig. A1(a) has been deteriorated, compared to the "cold" CTLs (blue symbols), and the modes no longer satisfy the relation $\Delta\omega \sim (\Delta k)^3$ coming from (1) but rather a perturbed version given in (8). However, for frequencies away from the SIP (not shown in Fig. A1) the modes asymptotically approach those of the cold structure because waves in the CTLs are no longer synchronized with the electron beam. Increasing the beam average current $I_0$ results in higher deformation of the dispersion diagram (red curves), and for a current of $I_0 = 0.5$ A the dispersion is so deformed that the original SIP of the "cold" structure cannot be recognized anymore. Interestingly (see for example the case of $I_0 = 50$ A), a feature that resembles the SIP is present in the hot structure as seen in Fig. A1(g). However, the SIP in the hot structure is at a slightly different frequency and wavenumber compared to the one in the cold SIP (blue symbols).

The degree of amplification in the interacting periodic SWS is characterized by the imaginary part of the Floquet-Bloch wavenumber for the modes having Re($k$) > 0 for an SWS with SIP interacting with an electron beam. We plot in Fig. A2 the scaling of the imaginary part of the Floquet-Bloch wavenumber with the beam current for the hot SWS-beam system, at the SIP angular frequency $\omega_{\text{SIP}}$ of the cold SWS. As explained there are five modes in Fig. A1 having Re($k$) > 0, however in Fig. A2 we plot only the three, out of these five, that have Im($k$) > 0, i.e., those that are amplified in their propagation along *z*. Thanks to the presence of the electron beam, the Floquet-Bloch wavenumber becomes complex for growing $I_0$ and it scales as $I_0^{1/3}$ as seen in Fig. A2 which signifies the strong amplification in the SWS for small beam current at SIP. Interestingly there is more than one eigenmode contributing to amplification, as was theorized in [31] though in this particular case one is dominant for large beam current and follows an $I_0^{1/3}$ trend. Note that for higher current (outside the range shown in Fig. A2) the $I_0^{1/3}$ trend is ceased because of the high effect of perturbations, however such large beam current scheme can be exploited when losses are present (as in gain and loss balanced systems, see [52], [53]).

APPENDIX B: SWS AND BEAM PARAMETERS USED IN SIMULATIONS

We consider a periodic multiple transmission line model composed of unit cells having three CTLs. Each unit cell has the period of $d = d_A + d_B = 25$ mm and is made of two consecutive segments *A* with length $d_A$ and *B* with length $d_B$, each made of three coupled TLs. The model is like the one in Sec. IV of [31] with the only difference that here the CTLs is made of 3 TLs while in [31], [32] the CTLs was made of only 2 TLs, though the matrix formalisms is the same. Therefore each 8×8 system matrix $\underline{\underline{\mathbf{M}}}_A$ and $\underline{\underline{\mathbf{M}}}_B$ in Sec. IV of [31] includes 3×3 matrices representing distributed inductance $\underline{\underline{\mathbf{L}}}_m$, distributed capacitance $\underline{\underline{\mathbf{C}}}_m$, and distributed series capacitance $\underline{\underline{\mathbf{C}}}_{m,c}$ (assumed diagonal), where *m* = *A,B*. The distributed series capacitance is used to model the high pass feature associated to the presence of cutoff frequencies in a waveguide, as shown in Ch. 8 of [42] and in [31], [32]. A distributed series resistance matrix $\underline{\underline{\mathbf{R}}}_m$ (assumed diagonal) is used to represent Ohmic losses in the metal.

The values are obtained based on a cylindrical metallic waveguide geometry supporting three coupled modes and are selected such that the average characteristic impedance (defined as the average of the characteristic impedances of all the segments of the three TLs in each unit cell) is around $50\,\Omega$ and develops an SIP around 1 GHz. Segment *A* of the CTLs has the length of $d_A$=10 mm and the following per unit length parameters: $L_{A1}$=0.56 µH/m, $L_{A2}$=0.07 µH/m, $L_{A3}$=0.74 µH/m, $C_{A1}$=80.6 pF/m, $C_{A2}$=58 pF/m, and $C_{A3}$=151.6 pF/m. For the coupling parameters of segment *A* we have $C_{A12}$=22 pF/m, and $C_{A23}$=0.7 pF/m. Segment *B* has the length of $d_B$=15 mm and the following per unit length parameters: $L_{B1}$=0.99 µH/m, $L_{B2}$=0.68 µH/m, $L_{B3}$=1.24 µH/m, $L_{B12}$=47 nH/m, $L_{B13}$=63 nH/m, $L_{B23}$=0.3 µH/m, $C_{B1}$=387 pF/m, $C_{B2}$=258 pF/m, $C_{B3}$=339 pF/m, and $C_{B12}$=25 pF/m, $C_{B23}$=111 pF/m and $C_{B13}$=145 pF/m. It should be noted here that in the matrix form of the per unit length capacitances, non-diagonal (coupling) elements have the negative values as in equation (3.24) of [54]. The distributed series capacitances per unit length are $C_{c1}$=0.6 pF·m, $C_{c2}$=0.5 pF·m, and $C_{c3}$=1 pF·m, in both CTLs segments *A* and *B*. SWS Ohmic losses in the metal are accounted in the CTLs as series per unit length resistances of 5 Ω/m, in each of the 3 TLs, in both segments *A* and *B*.

For the SWS with tilted-positive dispersion diagram only a few parameters are different compared to ideal SIP in segment *B*, and have the values of $L_{B12}$=78 nH/m, $L_{B13}$=1 nH/m. For the SWS with tilted-negative dispersion diagram the only different values are $L_{B12}$=240 nH/m, $L_{B13}$=310 nH/m and $C_{B13}$=97 pF/m. The length of the SWS interacting with the electron beam is $L=Nd$ where $N$ = 16 is the number of unit cells in every example, unless otherwise specified.

For the electron beam, in all the simulations we assume an average velocity of the electrons $u_0 = 0.3c$, therefore the beam voltage $V_0 = u_0^2/(2\eta) = 22{,}995\,\text{V}$, and we also assume the plasma angular frequency $\Omega_P = 0$ for simplicity. The threshold electron beam current to start oscillations in a SWS with ideal SIP is $I_{\text{th}} = 560\,\text{mA}$, whereas for a SWS with tilted-positive the starting electron beam current is $I_{\text{th}} = 880\,\text{mA}$. In both regimes of operations, the CTLs in Fig. 5 is terminated by $Z_{s1} = Z_{L1} = Z_{L2} = Z_{L3} = 75\,\Omega$, and the input terminals of the two lower TLs ($Z_{s2}$, $Z_{s3}$) are left open circuited. This loading has been used for all the results except for those in Figures 2, 3, 4 and 6.

The periodic single TL modeling a single mode TWT used for comparison is designed to have a unit cell with the same length as that of the SIP case (i.e., $d$ = 25 mm) and same average characteristic impedance ($50\,\Omega$) as the CTLs used for SIP schemes. Specifically, for comparison purposes we utilize a periodic TL since realistic TWTs are indeed constructed from periodic structures. This periodic single TL is composed of two segments *A* and *B* in each unit cell. The first segment has the physical length of $d_A$=10 mm and per unit length capacitance and inductance of $C_A$ = 56.41 pF/m and $L_A$ = 0.141 µH/m and the second segment has the physical length of $d_B$ =15 mm and per unit length capacitance and inductance of $C_B$ = 564.1 pF/m and $L_B$ = 1.41 µH/m. Also, we are assuming the cut-off series distributed capacitances of $C_c$ = 0.12 pF·m in each segment and a series per unit length resistances of 5Ω/m that represents losses. The starting electron beam current to start oscillations in a SWS made of a periodic single TL is $I_{\text{th}} = 4.7\,\text{A}$. Both ends of the single TL are terminated on $Z_s = Z_L = 75\,\Omega$ loads.